\documentclass[12pt,preprint]{aastex}
\usepackage{epsfig}
\usepackage{amssymb,amscd}
\usepackage{graphicx}
\usepackage{dcolumn}
\usepackage{bm}
\usepackage{natbib}

\begin{document}
\title{Quiescent times in Gamma-Ray-Bursts:\\ hints of a dormant inner engine}

\author{Alessandro Drago\altaffilmark{1,2} \& Giuseppe Pagliara\altaffilmark{3,2}}

\altaffiltext{1}{Dipartimento di Fisica, Universit{\`a} di Ferrara, 44100 Ferrara, Italy}
\altaffiltext{2}{INFN, Sezione di Ferrara, 44100 Ferrara, Italy}
\altaffiltext{3}{Dipartimento di Fisica, Politecnico di Torino, 10129 Torino, Italy}

\begin{abstract}

We perform a statistical analysis of the temporal
structure of long Gamma-Ray-Bursts (GRBs). 
First we consider a sample of bursts in which a long
quiescent time is present. Comparing
the pre-quiescence with the post-quiescence emission we show
that they display similar
temporal structures, hardness ratios and emitted powers, but,
on the average, the post-quiescence emission is roughly
twice as long as the pre-quiescence emission.
We then consider a sample of long and bright GRBs.
We show that the duration of each emission period
is compatible with the duration of an active 
period computed in various inner engine models.
At the contrary, if the inner engine is
assumed to be always active, i.e. also during the quiescent times, 
in several cases the total duration
of the burst largely exceeds the theoretical durations.
Our analysis therefore does not support the interpretation of long
quiescent times in terms of stochastic modulation of a continuous
wind. Instead the quiescent times can be interpreted as
dormancy periods of the inner engine. Before and after a dormancy period
the inner engine produces similar emissions.

\end{abstract}


\maketitle

\section{Introduction}
The time structure of GRBs is usually complex and
it often displays several short pulses separated by time intervals
lasting from fractions of second to several ten of seconds. The
analysis of the light curves can provide hints on the activity of the
inner engine although the relation between the observed signal and the
Physics of the inner engine is not yet completely understood.

A previous statistical analysis \citep{NP} has shown that there are
three time-scales in the GRB light curves: the shortest one is the
variability scale determining the pulses' durations and the intervals
between pulses; the largest one describes the total duration of the
bursts and, finally, an intermediate time scale is associated with long
periods within the bursts having no activity, the so called {\it
quiescent times}
\footnote{QTs are not the same as the
gaps between the precursors and the main pulses.  
In the case of the precursors, the gaps are between a softer
and weaker component (the precursor) and a harder and much stronger
one (the main burst). Here we observe gaps between pulses with the
same characteristics.}. The origin of these periods of quiescence is still unclear.

Here we show, through a statistical analysis, that if a quiescent time
longer than a few ten of seconds is present in the light curve then
the pre-quiescence and the post-quiescence emissions (PreQE, PostQE) have similar
variability scales but, on the average, the PostQE
is longer and only marginally softer than the PreQE. 
The similarities between the first and the second emission periods
strongly suggest that both emissions are produced by the same mechanism.
Moreover we will show that the average durations of PreQE and of PostQE, separately,
are compatible with the theoretical durations predicted by various inner enngine models.

\section{Data analysis}

\begin{figure}
\begin{center}
\includegraphics[scale=1]{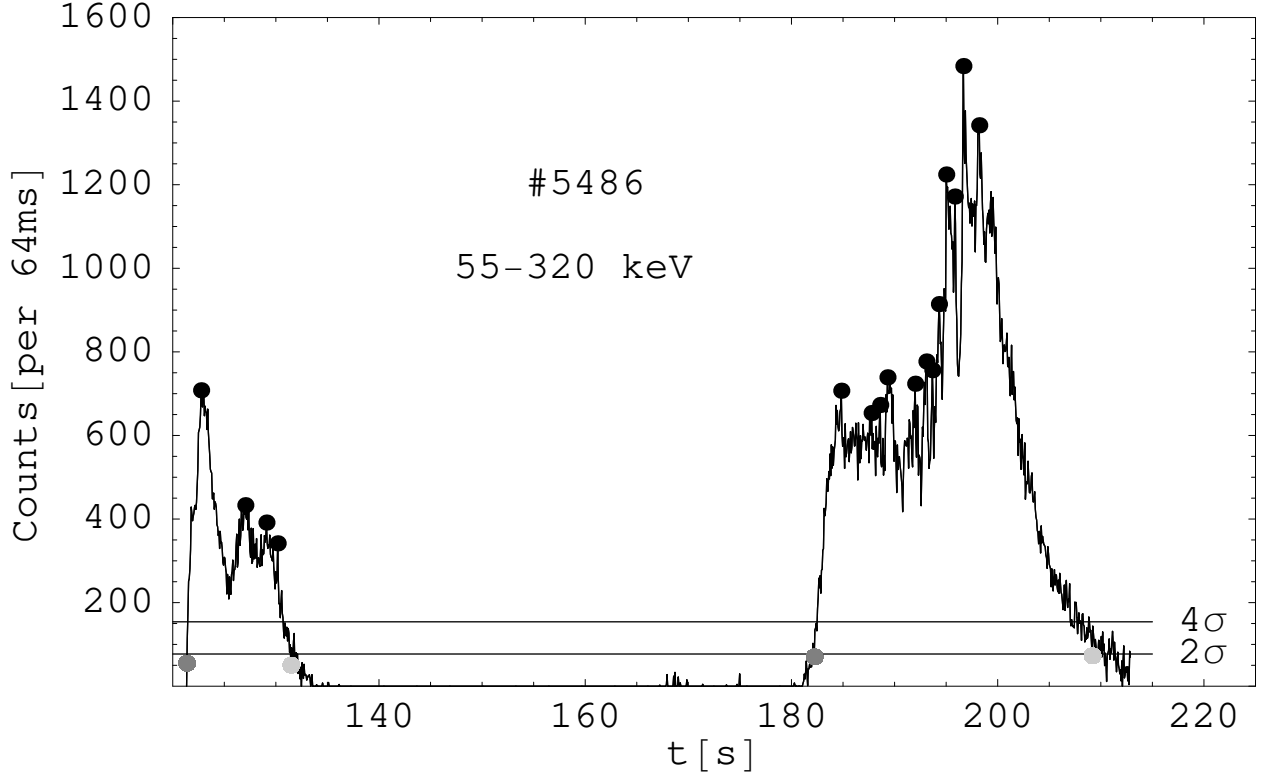}
\end{center}
\caption{{\bf  Light curve of a typical GRB} Time profile of BATSE
burst $\# 5486$ in the energy range $55 Kev <E< 320 Kev $. For each
light curve, the background is first determined using a linear fit and then
subtracted from the data. The signal is extracted taking as
initial and final bins the ones with counts above $2\sigma$.  
The active periods are defined as periods in which 
bins with counts exceeding 4$\sigma$ are present.
Active periods begin (dark-gray dots) and end (light-gray dots) when the signal 
drops below 2$\sigma$.
Within each active period we search for peaks using the peak
finding algorithm of \citet{Li-Fe}: indicating with $C_p$ the counts of a
candidate peak and with $C_1$ and $C_2$ the counts in the bins to the
left and to the right of the candidate, a ``true'' peak must satisfy the
relations $C_p -C_{1,2} \geq N_{var} \sqrt{C_p}$, where $N_{var} = 5$ in our
analysis. The true peaks are indicated by the black points.
\label{grb}}
\end{figure}

We have performed a statistical analysis of the time intervals $\Delta
t$ between adjacent peaks using the peak finding
algorithm of \citet{Li-Fe} and borrowing from \citet{NP1}
the definition of active periods separated by Quiescent Times (QTs).
In Fig.~\ref{grb} we show an example of burst where the previous
quantities are illustrated. We have applied this analysis to all the
light curves of the BATSE catalogue.

\begin{figure}
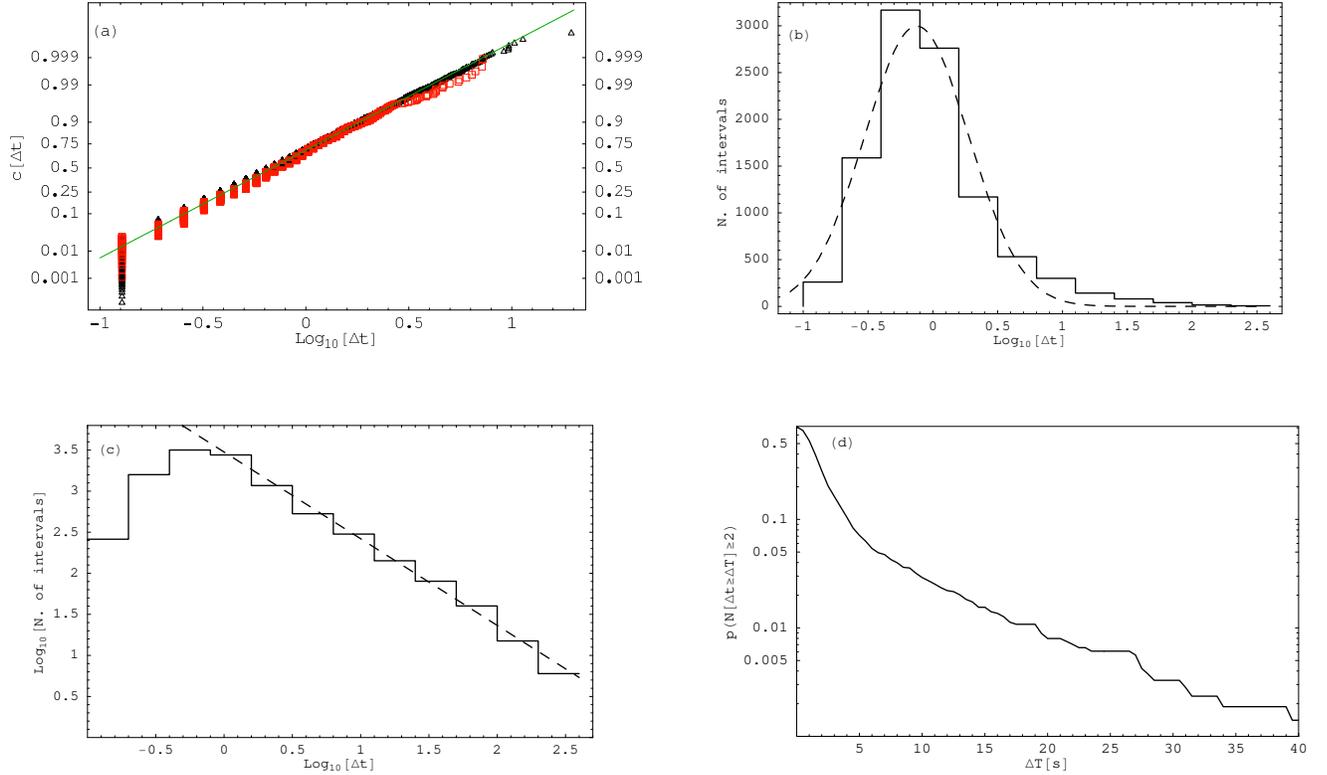

    \begin{centering}
\hbox{\hskip-0.cm \epsfig{file=f2a.epsi,height=4.6cm}\hskip 1.cm 
\epsfig{file=f2b.epsi,height=4.6cm}}
\vskip 1.cm    
\hbox{\hskip 0.3cm
\epsfig{file=f2c.epsi,height=4.6cm}
\hskip 1.6cm \epsfig{file=f2d.epsi,height=4.6cm}}
    \caption{{\bf  Analysis of time intervals between peaks} 
{\bf a} The cumulative distribution of time intervals $\Delta t$ which are not QTs
(black triangles), is compared with its best fit log-normal distribution (solid
green line). The data come from the full BATSE sample after
eliminating bursts displaying data gaps. The boxes correspond to
the cumulative distribution of $\Delta t$ taken from the subsample of
bursts containing a QT longer than $40$s (see text). {\bf b} Histogram
of the QTs and its log-normal fit (dashed line).
{\bf c} Histogram of QTs and power-low fit of its tail (dashed line).
The fit is based only on QTs longer than $40$s.
{\bf d} Frequency of bursts containing at least two QTs longer than $\Delta T$.
\label{rit} }
   \end{centering}
   \end{figure}

In a first investigation we have merged all the bursts of the
catalogue into one sample from which we compute the cumulative
probability $c(\Delta t)$ of finding time intervals $\Delta t$ which are not
QTs i.e.  we compute the distribution of the time intervals within each active period.  
In Fig.~\ref{rit}a, we show that $c(\Delta t)$ is well described by a
log-normal distribution. In Fig.~\ref{rit}b, the histogram of QTs is
displayed together with a log-normal distribution. As already observed
by previous authors \citep{NP}, there is an evident deviation of the
data points respect to the log-normal distribution for time intervals
longer than a few seconds, indicating an excess of long $\Delta t$.
In Fig.~\ref{rit}c we show a power law fit of the tail of the QTs
distribution which displays a very good agreement with the data, as
already observed by \citet{quilligan}.  The
physical interpretation of this distribution will be discussed later.
Finally, in Fig.~\ref{rit}d we show a correlation function, indicating
the probability of finding at least 2 QTs longer than $\Delta T$ in a
same GRB. As shown in the figure this probability rapidly decreases
and it essentially vanishes for $\Delta T > 40$ s.

A technical remark. In our analysis we have
considered two possible values for $\sigma$, the parameter used to
discriminate signal from noise.  A standard choice is
$\sigma=\sqrt{background}$.  Using that definition we can reproduce
the results of \citet{NP} on the cumulative distribution of $\Delta t$.  
Instead, the results presented in our figures have
been obtained using another
definition: $\sigma=\sqrt{\mathrm{Max}}$, where Max is the maximum number of
counts in the light curve. The reason for this choice is 
that we
want to study the main events of a GRB light curve (the ones
having a large luminosity) and not the faint
micro-structures (as for instance precursors, 
which we would exclude
from our sample).  It is important to remark that, using this larger
value of $\sigma$ we can recover the main results of 
Nakar and Piran, but the
distinction between a) time intervals which are not QTs and b) QTs 
is now even more clear on a
statistical basis: the firsts are perfectly interpreted by a
lognormal distribution, while long QTs are very well fitted by a
power-law, as discussed before.

We can now define a subsample of the BATSE catalogue composed of all
the bursts having a QT longer than 
$40$ s \footnote{The subsample contains the following 36 GRBs of BATSE catalogue:
142, 222, 869, 1328, 1989, 2138, 2148, 2156, 2211, 2213, 2922, 3336, 3351, 3488, 3634,
3776, 5421, 5478, 5486, 5585, 6295, 6335, 6454, 6472, 6629, 6745, 6892, 7170, 7185, 7301,
7503, 7549, 7769, 8001, 8063, 8087.}.  In Fig.~\ref{rit}a we also
show the distribution of $\Delta t$ within the subsample.  The
distributions of the full BATSE catalogue and of the subsample are
essentially equal.  This indicates that the
subsample is not composed of bursts having an anomalously
large redshift because instead all time scales within the subsample
would be homogeneously dilated.

In our analysis we will now concentrate on the subsample.  From the
result of Fig.~\ref{rit}d, the bursts of the subsample contain only one
long QT and it is therefore possible to divide each burst into a
PreQE and a PostQE of which
we will compare the temporal and spectral structure.

In Fig.~\ref{twoemission} we display the cumulative distributions
$c_1(\Delta t)$ and $c_2(\Delta t)$ within each of the two emission
periods. In panel a we display only $\Delta t$ which are not QTs 
(same as in Fig.~\ref{rit}a),
while in panel b QTs are included.
In both cases the two distributions are very similar. The $\chi^2$-test
provides a significance of $28\%$ for panel a and of $34\%$ for 
panel b that the two data sets are drawn
from the same distribution function \footnote{The $\chi^2$-test has been performed
on the set of $\Delta t$ after a binning and not on the cumulative distribution.
We have checked that using binnings in the range 2--8 s the result of
the test 
remains stable.}.  
Let us remind that within the
internal-external-shocks model \citep{piran2,meszaros}, external shocks produce
emissions lacking the short time scale variability produced by
internal shocks \citep{saripiran}. The result of Fig.~\ref{twoemission}
rules out a scenario in which PostQE is dominated by external shocks
and PreQE by internal shocks.  This in turn excludes the possibility
of associating the QTs with the time needed to the jet to reach and
interact with the interstellar medium. Clearly enough, the statistical
analysis we are presenting does not rule out the existence of specific
GRBs in which the second episode is indeed associated with external
shocks. For instance GRB960530 and GRB980125 are examples of bursts in
which PostQE has a smoother morphology and a
softer spectral evolution than PreQE \citep{hakkila}.

We perform now a statistical analysis of the durations $D1$ and $D2$
of the two emission periods.  As shown in Fig.~\ref{durations}a, the
two data sets are well fitted by two log-normal distributions (the
Kolmogorov-Smirnov test provides a significance of $\sim 90\%$ ).  The
two distributions have different mean values ($D1_{ave}\sim 21s$,
$D2_{ave}\sim 41s$) 
and almost
identical standard deviations ($\sigma_1=36s$, $\sigma_2=33s $). We
address now the following question: is the longer duration of PostQE a
manifestation of a progressive increase in the active periods'
durations during the burst?  To answer this question we have
repeated the previous analysis by dividing PreQE and PostQE each in
two parts, using the longest QT
within each emission as a divider.
The distributions of the duration of all parts are
shown in Fig.~\ref{durations}b. The durations of the two parts within
each emission period share the same distribution (the $\chi^2$-test provides
significances larger than $50\%$ in both cases) but, in agreement
with the previous findings, the average durations of the two parts of
PostQE are longer than the two parts of PreQE. Therefore, the longer
duration of PostQE cannot be attributed to a continuous modification
of the emission but is a specific feature of the second part of the
GRB.

To estimate the emitted energy during PreQE and PostQE we have
analyzed the hardness ratios, defined as the ratios between the
photon counts in two BATSE channels (the second and the third in our
case).  The average hardness of PostQE turns out to be only marginally
smaller ($\sim 20\%$) than the average hardness of PreQE.  
Since, as shown above, the average durations of PreQE and of PostQE
are in the relation $D2_{ave}\sim 2 \, D1_{ave}$, the total energy emitted during PostQE 
is also about a factor of 2 larger than the one emitted during PreQE.
This is also evident from the result of Fig.~\ref{power}, where the distributions of 
the powers emitted during PreQE and PostQE are displayed, showing that 
the average powers of the two emissions are essentially the same. 
 
It is interesting to compare our result on the durations with the correlation between
duration of QT and of PostQE found by \citet{ramireza}. By computing
the average duration of QT and of PostQE in our subsample (which is 
not the same analyzed by those authors)
we obtain the following result: PostQE $\sim$ 41 s and 
QT $\sim$ 80 s.  
Therefore, in our sample and using the Nakar and Piran
algorithm we cannot confirm the correlation found by \citet{ramireza}.
Indeed, if that correlation was present
the average durations of QT and of
PostQE should be comparable, since the correlation law suggested by
those authors is a straight line with a slope of order one.
Actually, in our sample the statistical correlation 
between the durations of QT and of PostQE is very small, $r \sim $ 0.13.

\section{Discussion}

\subsection{Dormant engine scenario {\it vs.} wind modulation model}

As observed by \citet{ramirez},
within the internal shocks model it is possible to explain the QTs
either as a turn-off of the IE or as a modulation of a
continuous relativistic wind emitted by the IE
(Wind Modulation Model WMM). Both
hypothesis are consistent with the result of
Fig.~\ref{twoemission}.

The main difference between the WMM and the dormant engine scenario
is that in the WMM the inner engine has to provide a constant
power during the whole duration of the burst. 
In our subsample, we have several bursts whose total duration
(including the QT) approaches 300 s. 
These durations have to be corrected taking into account the average
redshift of the BATSE catalogue, $z_{ave} \sim 2 $ \citep{piran}, but
even after this renormalization, durations of a hundred seconds or more are not
too rare. 
This time scale has to be compared with the typical duration of the
emission period of the inner engine, as estimated in various models.
For instance, in 
all numerical investigations of the collapsar model \citep{woosley,woosley1}
the IE remains switched-on
during some 20s.  
There is at the moment no indication that a 
"steady state disk", characterized by a mass infall rate 
large enough to power the GRB, can last hundreds of seconds instead
of tens of seconds. Also in the quark deconfinement model
\citep{cheng,datta,ouyed,apj,haensel}
the inner engine remains active during periods of the order of
a few ten seconds corresponding to the cooling time of the
compact stellar object.

To better clarify our argument concerning both the duration of an
emission period and the energy injected by the inner engine, let us
first discuss an example of a bright and long burst, GRB \#7301
of the BATSE catalogue (see Fig.~\ref{b2}). In
that burst a large QT is present lasting $\sim$ 100 s followed by a
PostQE of
$\sim$ 70 s and with a PreQE of $\sim$ 40 s. To compare these
time-scales to the typical durations provided by the models for the inner engine
we have first to divide all the observed durations by a factor of 3, due to
the average redshift of the BATSE catalogue.  While the resulting
durations of PreQE and of PostQE are then in agreement with the typical
time duration of an emission period in the collapsar model or in the
quark deconfinement model ($\sim$ 20 s), the total duration of the burst
($\sim$ 73 s, including the QT lasting $\sim$ 35 s), is roughly a factor of three larger
than the duration in the theoretical models. It is therefore
difficult to explain the long duration of the total burst within the
current inner engine models if the WMM is adopted.

Obviously, the previous remark would be meaningless if the
example we are providing is not typical of the time structure of
long GRBs. Instead, by checking all the GRBs of the 
"current catalogue" of BATSE, having T$_{90} >$ 100 s and peak photon flux
in the 256 ms channel $>$ 5 photons$^{-1}$cm$^{-2}$, we can come to
the following conclusion: all the lightcurves 
of this subsample can be interpreted within the dormant engine model
by rescaling
the total duration by a factor 1/(z+1) = 1/3 and by splitting the
burst in two emission episodes, separated by a 
dormancy period corresponding to a chosen QT.
In Figs.~\ref{b1},\ref{b2} we show all the GRBs of this sample
and we indicate the longest QT. 
Clearly enough, there are bursts
in which it is not unambiguous to decide which QT has to be interpreted as the
dormancy period, because more than one long QT is present. 
In the Figure we indicate the only two cases 
(out of 15 GRBs analyzed) in which
we suggest a dormancy period which is different from the
longest QT. In both those cases the suggested
dormancy period corresponds to the second longest QT present in the
burst. This rather small ambiguity can easily be explained by
noticing that a small fraction of long QTs can be generated by 
stochastic fluctuations (described by the tail of the 
log-normal distribution) and not by a dormancy
period. 
As discussed above,
the association of a QT with a dormancy period is
totally unambiguous for bursts in which a QT longer than
$\sim$ 40 s is present.
The last light curve presented in Fig.~\ref{b2} (GRB
$\#$ 6454) corresponds to a faint burst which does not
belong to the subsample discussed in this section
(GRBs of high luminosity).  It is
interesting to notice that in this burst PreQE and PostQE are so
long that also after dividing their durations by a redshift factor of
three, 
the durations are not compatible with the existing IE
models. Anyway, the faintness of this burst
is probably due to the extreme distance of the source and 
a larger red-shift correction should then be applied.
Similar considerations
are also valid for other superlong bursts presented in \citet{superlong}
(GRB $\#$ 6454 is one of the burst discussed in that paper). 
\footnote{
Using the algorithm of \citet{NP1} some 
faint components of the lightcurves are not considered as
parts of the signal because their counts number is smaller than 4-5
$\sigma$ and therefore they cannot be considered as active periods. 
This reflects on the estimated duration of the bursts.
In particular there are bursts whose $T_{90}$ durations exceed 100 s,
while their durations based on the active periods are shorter
and therefore these bursts are excluded from our sample.
This selection criteria clearly depends on the value of $\sigma$.
We have performed our analysis using for $\sigma$ both choices
described in Sec.~2. 
In Figs. 6-7 we show the results obtained using 
$\sigma=\sqrt{background}$, which is the most conservative choice
(in this case the excluded GRBs are $\#$ 1157, $\#$ 1886, $\#$ 3458, $\#$ 3930, $\#$ 7527). 
Using instead the larger value for $\sigma$, further 
components of the lightcurves would be excluded, and our 
conclusions would be even stronger.
}

Another problem with the WMM is due to the
prevision, within that model, that the emitted power during PostQE is
larger than during PreQE. This prevision is based again on
the existence of a continuous emission of shells also during the QT
\citep{ramirez}. The analysis displayed in Fig.~\ref{power} does
not support this prevision.

It is also worth recalling that, on the average, PostQE lasts roughly twice PreQE.
While it is probably possible to fix the parameters of
the WMM so to satisfy that constraint, no reason is
provided within the WMM model to explain that feature
in terms of the activity of the inner engine or
of the dynamics of the jet formation. 
At the contrary, models for a dormant inner engine can
associate the durations of the emission periods with
the durations of physical phenomena taking place within the
IE. This point will be discussed in the next Section.

We conclude that long QTs most probably correspond to periods of
inactivity of the IE.  On the other hand, it is possible
that the WMM is responsible for short QTs
occurring within the two emission periods, as suggested by
Fig.~\ref{durations}b where it is shown that the durations of the two parts of a single
emission period have the same distribution.
Finally, while our subsample is composed of GRBs having QTs longer
than 40 s, it is surely possible that in many cases the IE switches-off
for a much shorter time. Unfortunately, in those cases it is
not trivial to distinguish between QTs generated by the switch-off and
QTs generated by the WMM, as it also results from the discussion of 
Figs.~\ref{b1},\ref{b2}.

\subsection{Models for a dormant inner engine}

Let us now discuss how to generate dormancy periods using
various models of the IE.  Within the most
popular model, the collapsar model \citep{woosley}, there are two
possible scenarios: a temporary interruption of the jet produced by
Kelvin-Helmhotz instabilities \citep{woosley2} or the fragmentation of
the collapsing stellar core before its merging with the black hole \citep{osborne}. 
In both scenarios it can be possible to produce
long QTs.
For instance, in the scenario proposed by \citet{osborne}
the durations of the emission periods are related to the
durations of the accretion disks generated by each fragment.
Our result on the durations of PreQE and of PostQE can be explained
if, e.g.,
the average mass of the second fragment is larger than that of the 
first fragment. A discussion of the mass distribution of the fragments can be found 
in \citet{perna} \footnote{Disk fragmentation models could explain the 
recent X-ray-flare observations by Swift 
\citep{xray} which suggests the
possibility of a re-brightening of the IE.  In those cases,
however, the detected signal was softer than the GRB prompt emission
(the X-ray flares have not been detected by the BAT instrument on
Swift, but by the XRT). The re-brightening phenomena which we call
PostQEs belong instead to the prompt emission and they are only
marginally softer then the PreQEs. X-ray flares belong therefore to another class of phenomena.}.

Another model for the IE is based on the conversion of a
metastable hadronic star into a star containing quark
matter \citep{cheng,datta,ouyed,apj,haensel}. In the last years the 
possibility of forming
a diquark condensate at the center of a compact star has been widely
discussed in the literature \citep{raj}. The formation of a color
superconducting quark core can increase the energy released by a
significant amount \citep{noiprd}. It has also been shown that the
conversion from normal to gapped quark matter goes through a first
order transition \citep{shovk}. It is therefore tempting to associate
PreQE with the transition from hadronic to normal quark matter and
PostQE with the formation of the superconducting phase \citep{noi}.  In
this scenario the two dimensional scales regulating the durations of
PreQE and PostQE are the energies released in the two transitions.
Finally, let us remark that the power-law fitting the long QTs
distribution can originate from a superposition of exponential
distributions with different decay times \citep{flare}. In the quark
deconfinement model, the different decay times are associated with
slightly different masses of the metastable compact star.  
If the interpretation based on the quark deconfinement model
is correct then the GRBs data analysis provides very stringent bounds on the physics 
of high density matter.

A possible signature of the models in which the IE goes dormant
would be the detection of external shock emissions at the end of both 
PreQE and PostQE, indicating that the two emissions are physically
disconnected.

\section{Conclusions}

In this paper we have studied the problem of quiescent times in long GRBs
borrowing the technique developed by Nakar and Piran.
The main results obtained in our analysis are the following:

\noindent
--- the cumulative distribution of time intervals between peaks 
within a same active period is 
well fitted by a lognormal distribution, while
long quiescent times are well fitted by a powerlaw distribution.
This strengthens \citet{NP} conclusion about the different origin
of long quiescent times respect to shorter time intervals;

\noindent
--- the two components of the emission, preceding 
and following the quiescent time are statistically similar from the viewpoint of
their temporal microstructure, their hardness ratio and emitted power.
These results suggest a unique mechanism at the origin of both
the pre-quiescence and the post-quiescence emission.
Interestingly, the second emission lasts on the average twice the first,
what provides an important constraint to the inner engine models;

\noindent
--- the durations of the activity periods of the inner engine, computed within the existing 
theoretical models do not exceed few ten seconds. These theoretical estimates
compare favorably with the durations of the pre-quiescent and of the 
post-quiescent emissions, separately. On the other hand, 
in several bursts the total duration
including the quiescent time, is roughly three times longer. 
This is an indication in favor of a dormant inner engine 
respect to wind modulation models.

\section{Acknowledgments}

It is a pleasure to thank Filippo Frontera, Enrico Montanari and Rosalba Perna for many useful
suggestions and for help in the data analysis.

\begin{figure}
\begin{center}
\includegraphics[scale=1]{f3.epsi}
\end{center}
\caption{{\bf Analysis of time intervals between peaks within the two emission periods} 
{\bf a} The cumulative distribution of time intervals $\Delta t$ which are not QTs
are shown for the two emission episodes, PreQE and PostQE.
{\bf b} The cumulative distributions of $\Delta t$ are shown for the two emission episodes including QTs. 
\label{twoemission} }
\end{figure}

\begin{figure}
\begin{center}
\includegraphics[scale=1]{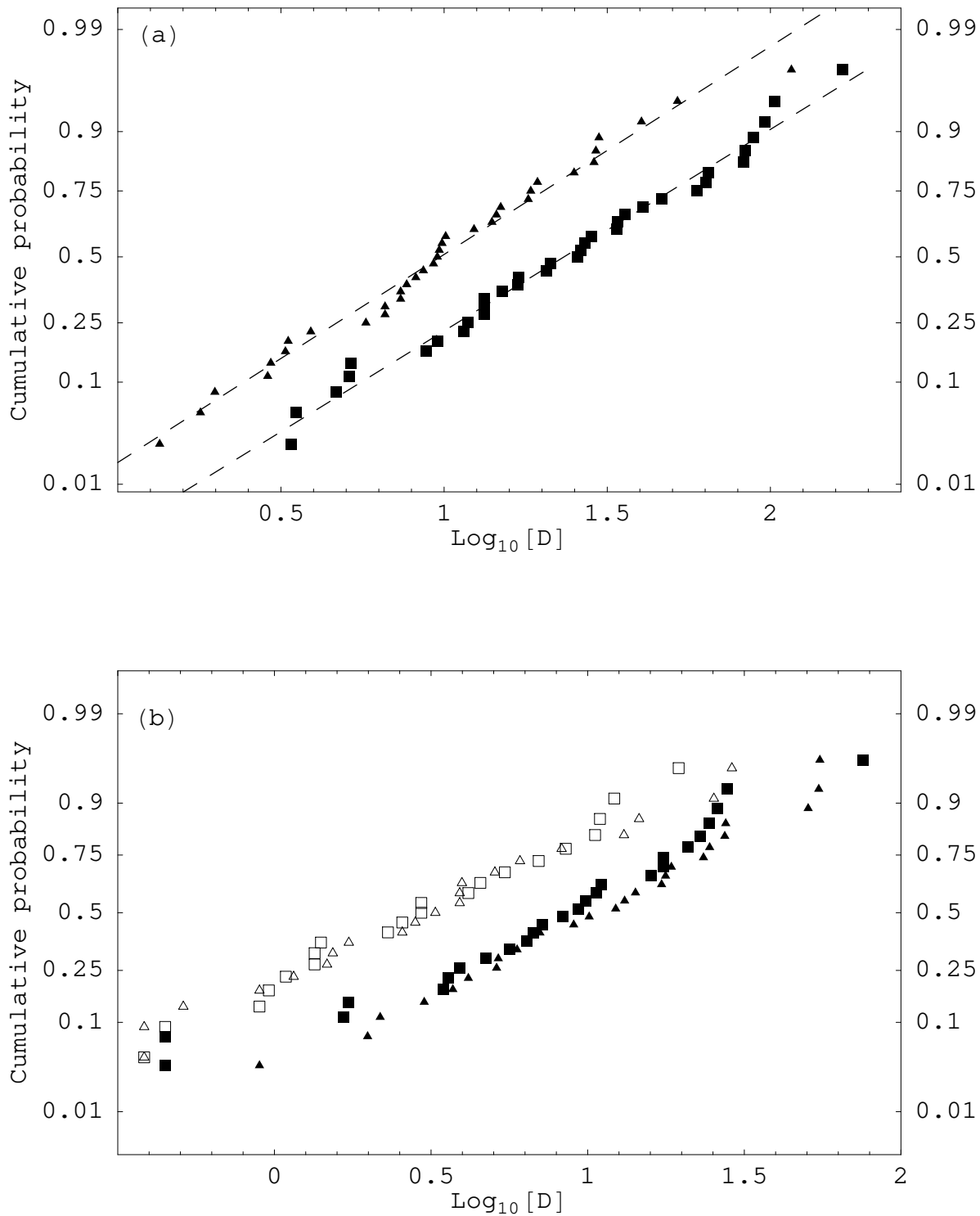}
\end{center}
\caption{{\bf Analysis of the durations $D$ of the two emission periods}
{\bf a} Cumulative distributions of durations of PreQE (filled triangles) and of PostQE
(filled boxes) and their best-fit log-normal distributions (dotted lines).
{\bf b} Cumulative distributions of durations of the first and second part 
of PreQE (empty triangles and boxes) and of PostQE (filled triangles and boxes).
\label{durations} }
\end{figure}

\begin{figure}
\begin{center}
\includegraphics[scale=1]{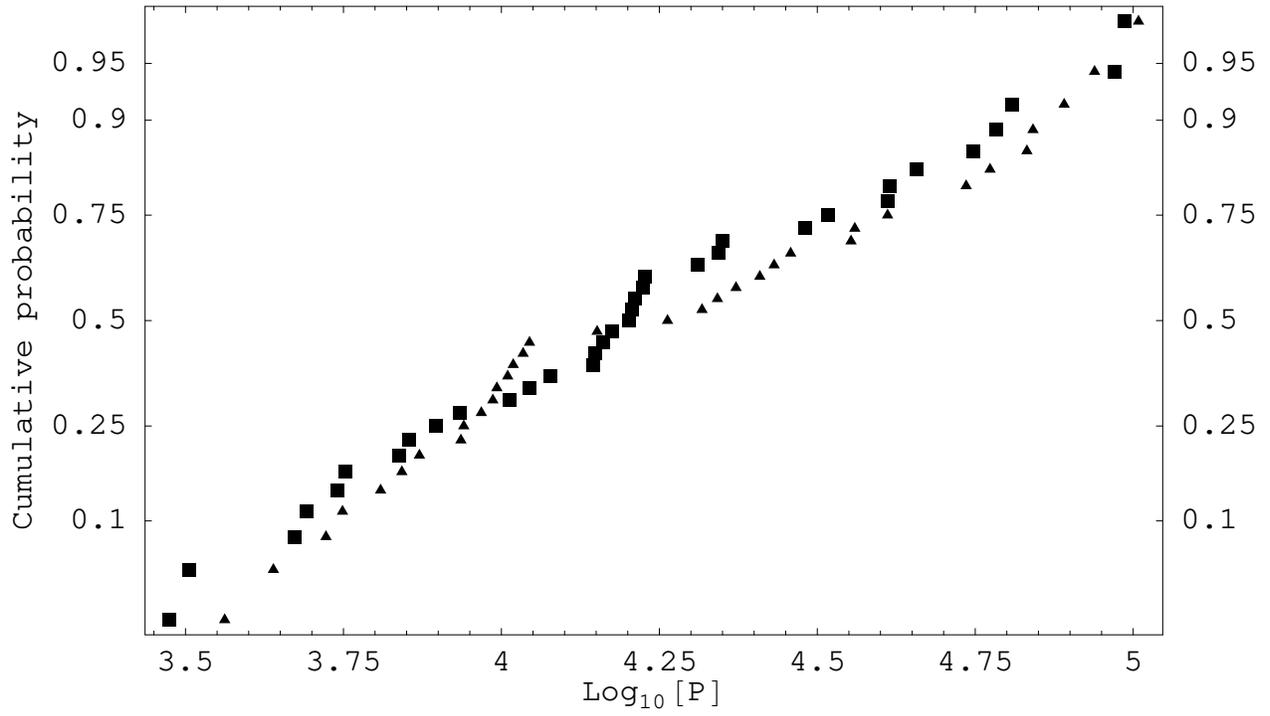}
\end{center}
\caption{{\bf Analysis of the powers $P$ of the two emission periods}
Cumulative distributions of powers of PreQE (filled triangles) and of PostQE (filled boxes)
\label{power} }
\end{figure}

\begin{figure}
\begin{center}
\includegraphics[scale=0.9]{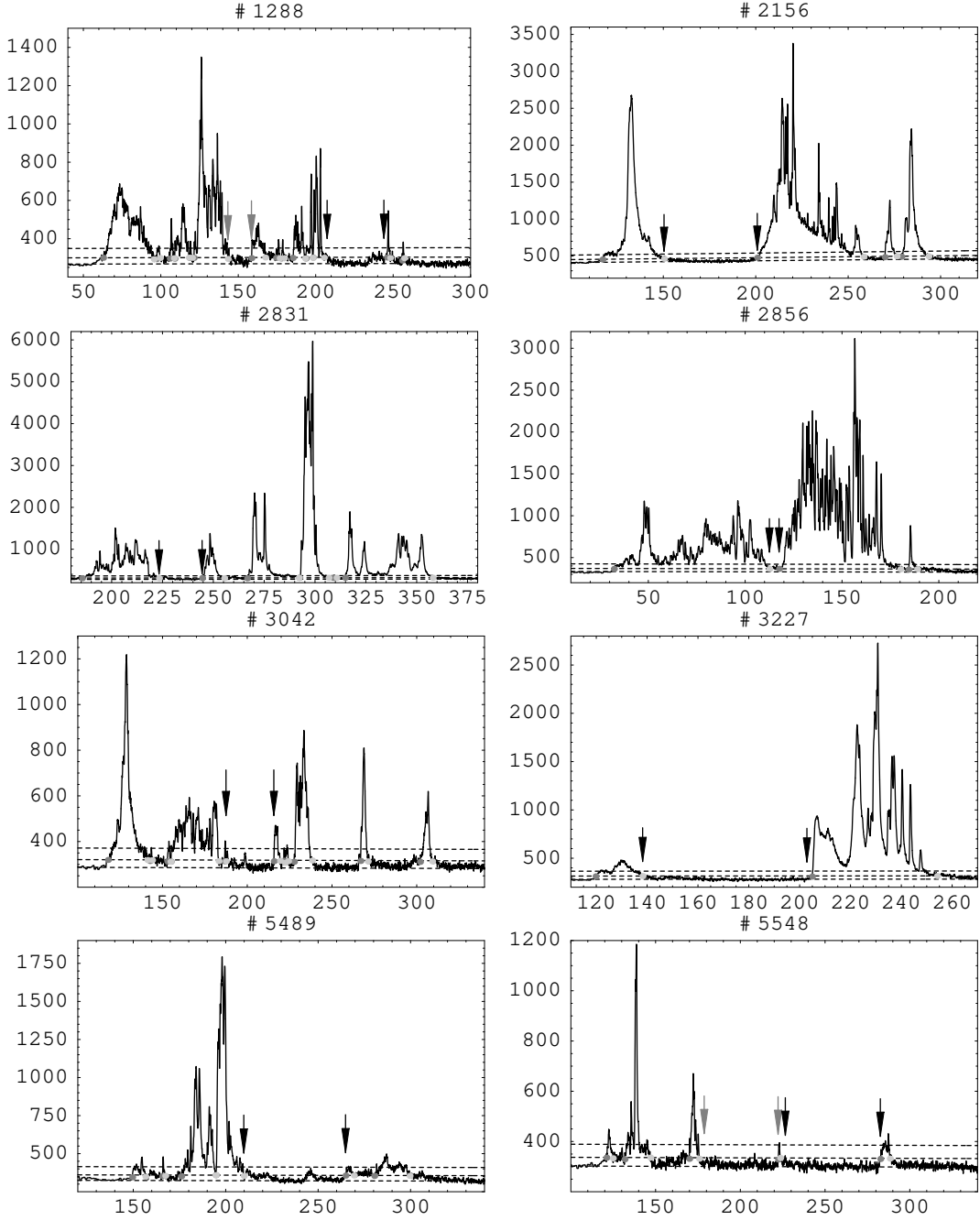}
\end{center}
\caption{{\bf Long and bright bursts  }
GRBs of the "current catalogue" of BATSE, having T$_{90} >$ 100 s and peak photon flux
in the 256 ms channel $>$ 5 photons$^{-1}$cm$^{-2}$. The dashed lines indicate the fit of
the background and the 2$\sigma$ and 5$\sigma$ levels used to define the active periods.
Dark-gray dots and light-gray dots mark the initial and the
end point of each active period, respectively.
Black arrows indicate the longest QT.
Gray arrows indicate the suggested dormancy period when
it does not coincide with the longest QT (two cases only).   
\label{b1} }
\end{figure}

\begin{figure}
\begin{center}
\includegraphics[scale=0.9]{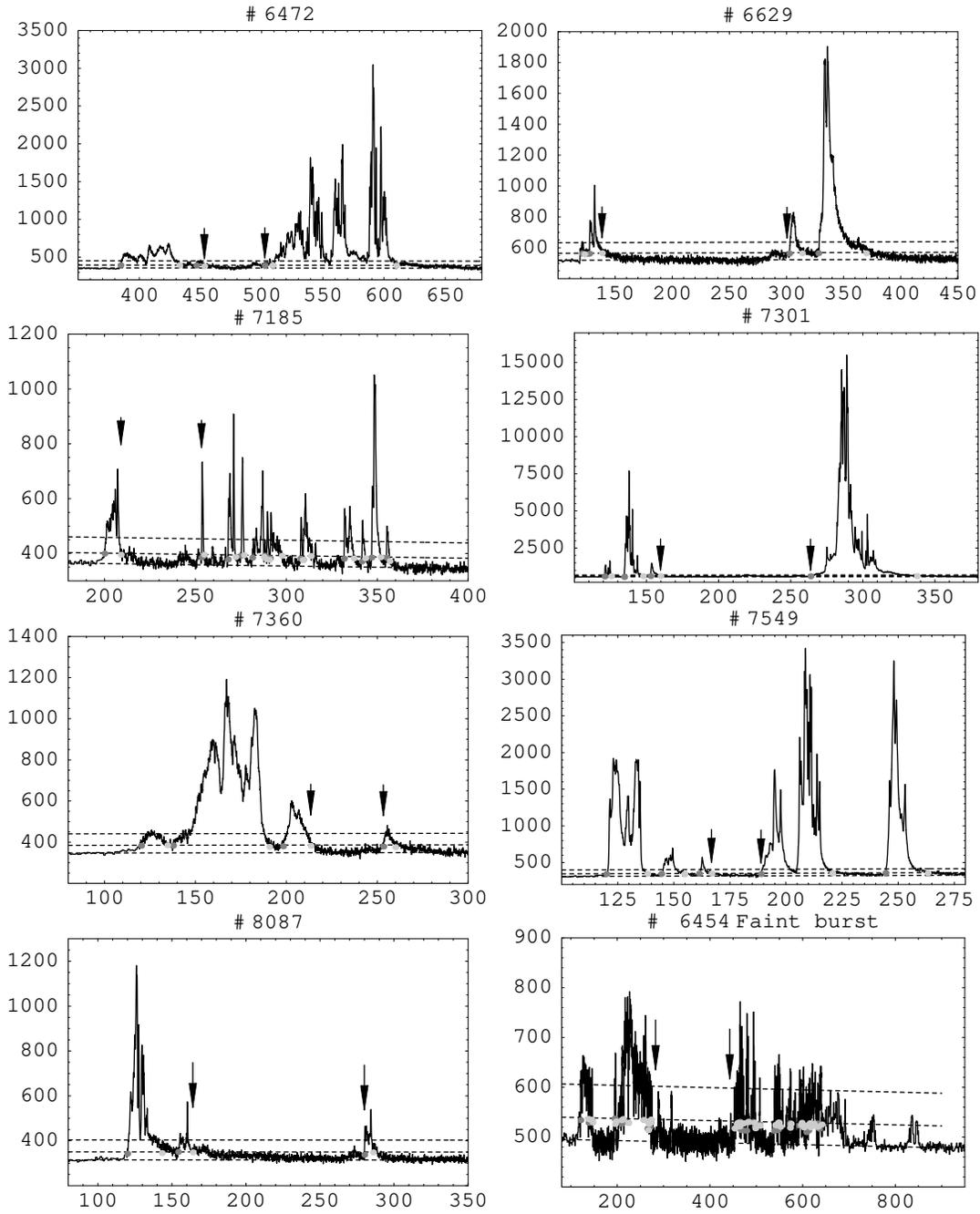}
\end{center}
\caption{{\bf Long and bright bursts}
Same as in Fig.~\ref{b1}. 
Here the suggested dormancy period
always coincide with the longest QT. We also display
a faint burst $\#$ 6454 (it does not satisfy the
limit on the peak photon flux) whose very long
duration could be due to a large red-shift.
\label{b2} }
\end{figure}

\end{document}